\documentstyle[11pt,newpasp,twoside,psfig]{article}
\def\iron{\hbox{[Fe\,{\sc ii}]}}

\def\he{\hbox{He\,{\sc i}}}
\def\sivi{\hbox{[Si\,{\sc vi}]}}
\def\sivii{\hbox{[Si\,{\sc vii}]}}
\def\caviii{\hbox{[Ca\,{\sc viii}]}}
\def\oiii{\hbox{[O\,{\sc iii}]}}

\def\oiii{\hbox{[O\,{\sc iii}]}}

\def\alix{\hbox{[Al\,{\sc ix}]}}

\def\brg{Br$\gamma$}

\def\h2{H$_2$}

\begin{document}

\title{Near--infrared spectroscopy of Seyfert galaxies}
\author{J.K. Kotilainen and J. Reunanen}
\affil{Tuorla Observatory, Piikki\"o, Finland}
\author{M.A. Prieto}
\affil{ESO, Garching, Germany}

\setcounter{page}{111}
\index{de Gaulle, C.}
\index{Churchill, W.}

\begin{abstract}
Results from near-infrared 1.5 -- 2.5 $\mu$m long-slit spectroscopy of 
14 nearby Seyfert galaxies are presented. 
\end{abstract}

\section{Introduction}

In the AGN unified models (Antonucci 1993), a molecular torus obscures the 
nucleus and the BLR in Seyfert 2 (S2) galaxies, while in Seyfert 1 
(S1) galaxies they are directly visible. Adequate spatial resolution to 
search for the molecular torus, with predicted size $<$ 100 pc 
(Pier \& Krolik 1993) can be achieved in the near-infrared (NIR). 
However, such NIR spectroscopic studies have so far been made at only 
moderate spatial resolution and/or only along the radio or {\oiii} axis 
(Veilleux et al. 1997; Winge et al. 2000). We present NIR 1.5 -- 2.5 $\mu$m 
long-slit spectroscopy of 14 nearby (0.002 $<$ z $<$ 0.010) Seyferts 
(3 S1s, 2 intermediate Seyferts and 9 S2s), both parallel and perpendicular 
to the ionization cone. For full discussion, see 
Reunanen. Kotilainen \& Prieto (2002a,b). 

\section{Results}

The strongest emission line is {\brg} 2.166 $\mu$m in 2 S1s, 2 intermediate 
Seyferts and 2 S2s; {\iron} 1.644 $\mu$m in 3 S2s; 
H$_2$ 1-0 S(1) 2.122 $\mu$m in 1 S1 and 1 S2; and a coronal line 
({\sivi} 1.964 $\mu$m or {\sivii} 2.483 $\mu$m) in 3 S2s.

Broad {\brg} is detected in 9 Seyferts, of which 5 are optically classified 
as S2s. The detection of broad {\brg} in more than half of the S2s is due to 
the lower nuclear extinction in the NIR. This extinction is unlikely to be 
related to the molecular torus, as its optical depth is likely to be high 
enough to obscure the BLR. The extinction is more likely caused by 
foreground dusty material in the form of star forming (SF) clouds, which 
frequently coexist with the AGN. 

Spatially resolved nuclear {\iron} emission was detected in 8 Seyferts (1 S1, 
2 intermediate Seyferts, and 5 S2s). The extended emission is patchy and 
follows closely the SF complexes as {\iron} is correlated with {\brg}. 
The nuclear {\iron} emission in a few Seyferts is consistent with 
X-ray excitation in the NLR, but appears predominantly to be shock-excited.  

The nuclear {\h2} surface density is higher in S2s than in S1s. This quantity 
is, however, correlated with the detection of broad \brg, indicating 
extinction effects. The {\h2} emission is extended in 11 Seyferts. In a few 
Seyferts (NGC 1097, NGC 1386, NGC 4945) off-nuclear {\h2} emission 
regions are detected, associated with the SF rings, but in most Seyferts, 
the extended molecular disc-like gas declines smoothly with radius. 
The spatial extent of the nuclear {\h2} emission is larger perpendicular to 
the cone than parallel to it in only 6/11 galaxies, in only moderate 
agreement with the unified models and the existence of a molecular torus. 
Thus a larger sample and/or higher spatial resolution data is required to 
search for the molecular material associated with the torus. Only in 
5 Seyferts, nuclear H$_2$ 2--1 S(1) 2.248 $\mu$m emission was detected and 
the 2--1 S(1)/1--0 S(1) ratios are inconsistent with significant 
fluorescent excitation, and agree with thermal excitation by X-rays or, 
more likely, by shocks.

Four coronal lines were detected: {\sivi}, {\alix} 2.043 $\mu$m, 
{\caviii} 2.321 $\mu$m and {\sivii}. Of these, {\sivi} was detected in 7, 
{\alix} in 3, {\caviii} in 4 and {\sivii} in 7 Seyferts. At least one 
coronal line was detected in 8 Seyferts (1 S1, 1 intermediate Seyfert and 
6 S2s), substantially increasing the number of Seyferts with NIR 
coronal line detection. Interestingly, in all three Seyferts with 
spatial information (NGC 1068, NGC 3081 and ESO 428-G14), the coronal line 
emission is extended parallel to the cone, indicating an anisotropic 
radiation field. Due to the high ionization potential of the coronal lines, 
the extended emission is mainly produced by shocks interacting with the 
interstellar medium.

\section{Work in progress: stellar populations}

The NIR spectra contain many stellar and nebular features 
({\he} 2.058 $\mu$m and 
{\brg} emission lines, Si I 1.588 $\mu$m and CO (6-3) 1.619 $\mu$m 
absorption lines, and the $^{12}$CO and $^{13}$CO absorption bandheads 
longward of 2.29 $\mu$m). The {\he} and {\brg} emission lines are related to 
the number of ionizing UV photons from massive stars, while the CO lines 
measure the stellar kinematics, light-to-mass ratio, and recent SF 
(Oliva et al. 1995). In Seyferts, the central CO emission is often diluted by 
hot dust emission heated by the AGN. 

We are currently studying the stellar population content of the Seyferts 
(Reunanen et al., in prep.). Stellar emission and absorption features probe 
the age and SF properties of the stellar population in the nucleus and as a 
function of radius. Comparison of these properties between Seyferts and a 
matched sample of spirals, for which we have obtained similar NIR spectra, 
will assess any relationship of SF with the nuclear power.


\begin{references}
\reference Antonucci R., 1993, ARA\&A 31, 473
\reference Oliva,E., Origlia,L., Kotilainen,J.K., Moorwood,A.F.M., 1995, A\&A 301, 55 
\reference Pier,E.A., Krolik,J.H., 1993 ApJ 418, 673
\reference Reunanen J., Kotilainen J.K., Prieto M.A. 2002a, MNRAS, 331, 154
\reference Reunanen J., Kotilainen J.K., Prieto M.A. 2002b, MNRAS, submitted
\reference Veilleux S., Goodrich R.W., Hill G.J. 1997, ApJ, 477, 631
\reference Winge C. et al. 2000, MNRAS, 316, 1 

\end{references}
\end{document}